\documentclass[aps,10pt,prd,twocolumn,preprintnumbers,amsmath,amssymb,nofootinbib,superscriptaddress,a4paper,showpacs]{revtex4-1}

\usepackage[breaklinks]{hyperref}
\usepackage[utf8]{inputenc}
\usepackage{graphicx}
\usepackage{color}
\usepackage{amsfonts,amsthm}
\usepackage{bm,bbm}

\newcommand{\be}{\begin{equation}}
\newcommand{\ee}{\end{equation}}
\newcommand{\ba}{\begin{eqnarray}}
\newcommand{\ea}{\end{eqnarray}}
\def\bs{\begin{subequations}}
\def\es{\end{subequations}}

\def\a{\alpha}
\def\b{\beta}

\def\la{\lambda}

\def\s{\sigma}

\def\cC{\mathcal{C}}

\def\cH{\mathcal{H}}

\def\cL{\mathcal{L}}

\def\cN{\mathcal{N}}

\def\cP{\mathcal{P}}

\def\cV{\mathcal{V}}

\def\ds{d_{\rm S}}
\def\dh{d_{\rm H}}

\def\p{\partial}

\newcommand{\Eq}[1]{(\ref{#1})}
\def\com{\color{magenta}}
\def\cob{\color{blue}}
\newcommand{\oarX}[1]{\href{http://arxiv.org/abs/#1}{{\ttfamily\com arXiv:#1}}}
\newcommand{\arX}[1]{\href{http://arxiv.org/abs/#1}{{\ttfamily\com arXiv:#1}}}
\newcommand{\doin}[6]{\href{http://dx.doi.org/#1}{{\cob #2 #3 {\bf #4}, #5 (#6)}}}
\newcommand{\doinn}[5]{\href{http://dx.doi.org/#1}{{\cob #2 {\bf #3}, #4 (#5)}}}
\newcommand{\doij}[5]{\href{http://dx.doi.org/#1}{{\cob #2 #3 (#5) #4}}}

\newcommand{\proc}[6]{in \emph{#1}, edited by #2 (#3, #4, #5, #6)}
\newcommand{\procsin}[5]{in \emph{#1}, edited by #2 (#3, #4, #5)}
\newcommand{\tia}[1]{}

\def\lp{\ell_{\rm Pl}}

\def\rme{e}
\def\rmd{d}
\def\rmi{i}

\begin{document}

\title{Deformed symmetries in noncommutative and multifractional spacetimes}

\author{Gianluca Calcagni}
\email{calcagni@iem.cfmac.csic.es}
\affiliation{Instituto de Estructura de la Materia, CSIC, Serrano 121, 28006 Madrid, Spain}

\author{Michele Ronco}
\email{michele.ronco@roma1.infn.it}
\affiliation{Dipartimento di Fisica, Universit\`a di Roma ``La Sapienza", Piazzale Aldo Moro 2, 00185 Roma, Italy}
\affiliation{INFN, Sezione Roma1, P.le A. Moro 2, 00185 Roma, Italy}

\date{August 4, 2016}

\begin{abstract}
We clarify the relation between noncommutative spacetimes and multifractional geometries, two quantum-gravity-related approaches where the fundamental description of spacetime is not given by a classical smooth geometry. Despite their different conceptual premises and mathematical formalisms, both research programs allow for the spacetime dimension to vary with the probed scale. This feature and other similarities led to ask whether there is a duality between these two independent proposals. In the absence of curvature and comparing the symmetries of both position and momentum space, we show that $\kappa$-Minkowski spacetime and the commutative multifractional theory with $q$-derivatives are physically inequivalent but they admit several contact points that allow one to describe certain aspects of  $\kappa$-Minkowski noncommutative geometry as a multifractional theory and vice versa. Contrary to previous literature, this result holds without assuming any specific measure for $\kappa$-Minkowski. More generally, no well-defined $\star$-product can be constructed from the $q$-theory, although the latter does admit a natural noncommutative extension with a given deformed Poincar\'e algebra. A similar no-go theorem may be valid for all multiscale theories with factorizable measures. Turning gravity on, we write the algebras of gravitational first-class constraints in the multifractional theories with $q$- and weighted derivatives and discuss their differences with respect to the deformed algebras of $\kappa$-Minkowski spacetime and of loop quantum gravity. 
\end{abstract}

\preprint{\doin{10.1103/PhysRevD.95.045001}{PHYSICAL REVIEW}{D}{95}{045001}{2017} \hspace{9cm} \arX{1608.01667}}



\maketitle


\section{Introduction}

The deformation of the symmetries of general relativity is a typical feature of quantum-gravity scenarios. Effects of quantum or anomalous geometry can break Poincar\'e symmetries in local inertial frames as well as diffeomorphisms at a global level. The theory may still be invariant under other types of symmetries, which typically are a deformation of classical Poincar\'e and diffeomorphism symmetries. Thus, there are two meanings in which one has a \emph{deformed algebra} of the generators of such symmetries. One is by deforming the generators $A_i\to A_i'$, which corresponds to a deformation of classical symmetries. For instance, in the quantum theory one can have a momentum operator $P_i'$ which generates a symmetry $x_i\to f(x_i)$ analogous to the usual spatial translations $x_i\to x_i+a_i$ generated by $P_i$, such that $f(x_i)\simeq x_i+a_i$ when quantum corrections are negligible. In this case, one regards $P_i'$ as the generator of ``deformed spatial translations.'' The other way in which an algebra is deformed is by a change in its structure. For instance, given a classical algebra $\{A_i,A_j\}=f_{ij}^kA_k$ one might end up with an algebra $\{A_i',A_j'\}=F(A_k')$ in the quantum theory, which can be written also in terms of the generators of the classical symmetries, $\{A_i,A_j\}=G(A_k)$, for some $G\neq F$.

Given the proliferation of theories where spacetime geometry is heavily deformed by quantum-gravity effects or by other mechanisms, it is important to discriminate among all the different ways in which symmetry algebras are deformed. This task is all the more urgent considering that many such theories share some striking similarities. For instance, all of them are characterized by dimensional flow, the change of spacetime dimensionality with the probed scale \cite{tH93,Car09,fra1}. Three examples, which are the focus of the present paper, are noncommutative spacetimes \cite{Ben08,AA,ArTr1,AnHa1,AnHa2}, multiscale (in particular, multifractional) spacetimes with weighted and $q$-derivatives \cite{frc2,frc7} and loop quantum gravity \cite{COT2,COT3}. We recall that \emph{multiscale} theories are such that the effective dimensions of spacetime, suitably defined, change with the observation scale (dimensional flow). \emph{Multifractional} spacetimes are multiscale spacetimes whose measures in position and momentum space are factorizable in the coordinates.

Noncommutative and multifractional spacetimes capture, in different manners, some features that, according to evidence, expectations, or intuition, should characterize quantum gravity. In particular, spacetime noncommutativity aims to give a picture of a quantum Minkowski spacetime by promoting coordinates to noncommutative operators. Independent studies have shown that it can arise from full-fledged theories of quantum gravity \cite{CoDS,ChHo,Sch99,SeWi,ACSS,GLO1,FrLi2,BaOr1,BDOT,ACFKS,hdaPa}, thereby enforcing the belief that quantum gravity might require a description in terms of spacetime noncommutativity in its flat regime. More importantly, it is possible to extract effective physics and make predictions, some of which have been of direct relevance for phenomenology \cite{AEMNS,Ame98,ACMa}. Among them we remind energy-dependent dispersion of radiation, in-vacuo birefringence, modified composition laws for momenta, blurring images of distant sources due to spacetime fuzziness, and so on \cite{Ame08,Mat05}. One of the main motivations for studying multifractional geometries comes from the fact that they are simple models to realize dimensional flow, a feature common to all quantum-gravity approaches and in some cases related to improved renormalization properties \cite{revmu}. They explore the possibility that spacetime is not a continuous manifold but, rather, a multiscale geometry or, in some cases, even a multifractal. To account for a fractal structure, the integration measure and kinetic terms have to be deformed. Fractal properties of geometry may arise in an intermediate regime between the infrared continuous spacetime that we perceive and a discrete fundamental picture (available in some other top-down quantum-gravity approaches) at the ultraviolet scale. Remarkably, the hypothesis of fractality gives exactly the same measure required, more generally, when the infrared is reached as an asymptote, and it produces a very rich phenomenology that goes from effects on electroweak processes to gravitational waves, to modifications of the cosmic microwave background spectra, and more \cite{revmu}. Multifractional frameworks are independent theories with a top-down construction, but they can also be regarded as bottom-up effective models in certain regimes of interest for phenomenology \cite{revmu}. In parallel to these approaches, loop quantum gravity could, in principle, provide a full solution to the problem of quantizing the gravitational interaction; its phenomenology is still under construction.

Experimental data are the best (and, of course, a necessary) guidance in the construction of physical theories. This is the main reason why approaches such as noncommutative spacetimes and multifractional geometries, with so many contacts with  phenomenology, have been developed and will not easily get old. However, the search for convergences or, on the opposite, departures between different approaches which are attracting interest in the literature, represent one of the ways to make some progress in quantum gravity. We can benefit from the study of links or dualities among quantum-gravity proposals for several reasons. First of all, dualities can clear up the highly diversified panorama of quantum-gravity research and, moreover, yield new insights and bring novel physical predictions. Bottom-up approaches are, by construction, incomplete and each of them addresses the quantum-gravity problem from a different perspective. Then, it is natural to regard them as different pieces of the same jigsaw puzzle and to look for the existence of possible complementarities. Furthermore, due to the fact that they focus on different features, bottom-up approaches use very different formalisms designed to develop a certain characteristic of quantum gravity. As a consequence, they may be useful and powerful in the characterization of some aspects, but weak or inadequate in the description of others. For instance, while dimensional flow is the cornerstone of multifractional geometries, there are fragmentary hints of its presence in noncommutative spacetimes and, in certain cases, it remains difficult to derive it rigorously. On the other hand, integration measures can be either factorizable or not in noncommutative spacetimes, but we do not consider nonfactorizable measures in the context of multifractional geometries. In the light of this, finding that these approaches are complementary would contribute positively to their development. Finally, it is important to look for synergies between top-down and bottom-up approaches because the latter are not candidate quantum-gravity theories (they always rest on some kind of simplification) and they need to be embedded into a top-down proposal. Conversely, the complexity of top-down approaches often forbids us to extract physical predictions.

Some relations among these frameworks have been explored in the past. In Ref.\ \cite{ACOS}, it was shown that the cyclicity-inducing measure of $\kappa$-Minkowski spacetime can be reproduced by the spacetime measure of multifractional theories in the limit of very small scales. This suggested a tight relation, or even a duality, between $\kappa$-Minkowski spacetime and some multifractional theory. However, in order to have a duality it remained to show that both theories have the same symmetries. Since the publication of Ref.\ \cite{ACOS}, this issue remained unaddressed. We feel that the relation between noncommutative spacetimes and multifractional geometries deserves to be further studied and fully clarified. For the aforementioned reasons, determining what is such relation is important not only from a mathematical point of view, but also and especially from the perspective of phenomenology. The main objective of this work is to perform a complete and self-contained analysis of this problem. In this paper, we will fill this gap and conclude that, although $\kappa$-Minkowski is not exactly dual to any of the known multifractional theories, it shares a number of similarities which permit to describe, in certain regimes, this noncommutative spacetime as a multifractional one and vice versa.  

In the process, we will recover previous results in a more general way. In Ref.\ \cite{ACOS}, a class of noncommutative spacetimes was constructed such that their cyclicity-inducing measures in position space coincide, after inspecting the Heisenberg algebra of spacetime coordinates, with a specific fractional measure $\sim x^\a$ employed in multifractional theories. Contrary to that approach, we will face this problem at the level of the Poincar\'e algebra and find a correspondence between $\kappa$-Minkowski and the noncommutative version of a certain multifractional spacetime, \emph{without imposing cyclicity invariance}. Generalizing to an arbitrary multifractional measure, we will obtain a class of noncommutative spacetimes endowed with a certain deformed Poincar\'e algebra, which we will write down explicitly.

Note that noncommutative spacetimes do have dimensional flow \cite{Ben08,AA,ArTr1} and, therefore, are multiscale by definition \cite{trtls}. The issue here is whether they are dual to commutative multifractional spacetimes, which are a special case of multiscale geometries. We immediately spell out the main reason why one cannot establish an exact duality between $\kappa$-Minkowski and any of the commutative multifractional theories: multifractional measures are always factorizable both in position and in momentum space, while, in general, the measures of $\kappa$-Minkowski in position and momentum space do not enjoy this property. It is therefore natural to find different symmetries in these theories. These findings lead us to a reconsideration of the mutual standing of noncommutative and multifractional theories: rather then being dual to each other, they are one the extension of the other to the case of nonfactorizable position or momentum measures. They simply cover different regions in the landscape of multiscale theories (roughly sketched in \cite{trtls}).

In parallel, a connection between $\kappa$-Minkowski spacetime and the effective-dynamics (or effective-constraint, or deformed-algebra) approach of loop quantum gravity (LQG) \cite{bojopaily,bojopaily1,BBCGK} was found recently \cite{hdaPa}: the deformed Poincaré symmetries of the two theories are mutually compatible. To see this, one plugs the Killing vectors of Minkowski spacetime into the LQG deformed constraint algebra and recasts the first-class constrains in terms of rotation, boost and translation operators on flat space. Since there is a relation between $\kappa$-Minkowski and multifractional spacetimes, one may wonder if there is also a relation between the latter and the effective limit of loop quantum gravity described by the deformed-algebra approach. Such relation will not be a duality for the reasons explained above: if the symmetries of $\kappa$-Minkowski are compatible with those of loop quantum gravity but different from the symmetries of multifractional spacetimes, then the latter cannot be exactly equivalent to loop quantum gravity. Nevertheless, it is possible to construct the deformed algebra of the gravitational constraints in two multifractional theories (with $q$- or weighted derivatives) and compare it directly with the anomaly-free algebra found in the effective-dynamics approach of LQG. We will do so here and discuss similarities and differences in the deformations.

The plan of the paper is as follows. In Sec.\ \ref{2}, we explore the possibility to interpret multifractional spacetimes as noncommutative spacetimes in disguise. In Sec.\ \ref{2b}, we take the bicross-product Casimir operator in $\kappa$-Minkowski and find the corresponding multifractional measure in position space; however, the corresponding momentum measure is different from that of $\kappa$-Minkowski. In Sec.\ \ref{2c}, we turn the problem around and look for noncommutative spacetimes dual to the multifractional theory with $q$-derivatives. Interpreting the product of two multifractional plane waves as the $\star$-product of two normal phases, one can extract an approximate commutation relation of the coordinates. However, the latter is ill-defined because it features the momenta of both plane waves. The cause is identified in the factorizability of the multifractional measure. In Sec.\ \ref{3}, we move our focus onto the type of noncommutative algebras arising when the coordinates in position and momentum space of multifractional theories are promoted to noncommuting operators. In Sec.\ \ref{3a}, we rederive the main result of \cite{ACOS} with a much easier and faster method which also has the virtue of being independent of the requirement of cyclicity invariance. Instead of applying the Weyl map on plane waves as done in \cite{ACOS}, we will simply compute the phase-space (position-momentum) Heisenberg algebra. Thus, the connection found in \cite{ACOS} between $\kappa$-Minkowski and multifractional spacetimes turns out to be much more general than in its original derivation. These results are generalized, in Sec.\ \ref{3b}, to arbitrary measures as well as Poincar\'e and Heisenberg algebras. A comparison between the deformed gravity algebra of LQG and the one in the multifractional theories with $q$- and weighted derivatives is carried out in Sec.\ \ref{4}. Section \ref{concl} is devoted to conclusions.

In most of the paper, we work in $1+1$ dimensions for simplicity and with time-space signature $(-,+)$. The generalization to many spatial directions is straightforward and does not entail any relevant modification of the results presented here. $\hbar=1=c$ throughout and $8\pi G=1$ in Sec.\ \ref{4}.


\section{A no-go theorem: separation in the multiscale landscape}\label{2}


\subsection{Essential facts about multifractional theories}

The purpose of this subsection is to introduce only the technical ingredients of multifractional spacetimes needed in the paper; it is not meant to give a self-contained, exhaustive introduction on the subject. The reader is encouraged to consult the bibliography for all details concerning theoretical foundations, conceptual topics, physical interpretation and phenomenology. Recent overview sections can be found in \cite{trtls,CKT}.

The first element we will use is the existence of a factorizable nontrivial measure in position and in momentum space. By definition, any given multifractional field-theory action $S=\int \rmd^Dq(x)\,\cL$ in $D$ topological dimensions is characterized by a measure $\rmd^Dq(x)=\rmd q^0(x^0)\,\rmd q^1(x^1)\cdots \rmd q^{D-1}(x^{D-1})=\rmd^Dx\,v_0(x^0)\cdots v_{D-1}(x^{D-1})$, where $q^\mu(x^\mu)$ are called geometric coordinates and $v_\mu(x^\mu)>0$ are $D$ measure weights, possibly all different from one another. The symmetries of the Lagrangian $\cL$ depend on the choice of kinetic terms for the field. In turn, these symmetries determine the measure $\rmd^Dp(k)=\rmd p^0(k^0)\cdots \rmd p^{D-1}(k^{D-1})=\rmd^Dk\,w_0(k^0)\cdots w_{D-1}(k^{D-1})$ in momentum space. Of the four extant multifractional theories, we will consider only those with so-called $q$-derivatives and with weighted derivatives. The former, where all derivative operators $\p_\mu=\p/\p x^\mu$ in the field-theory action are replaced by $\p/\p q^\mu(x^\mu)$ (called $q$-derivatives), is characterized by a specific relation between position and momentum geometric coordinates, which are canonically conjugate variables \cite{frc11}:
\be\label{pqmulti}
p^\mu(k^\mu)=\frac{1}{q^\mu(1/k^\mu)}\,.
\ee
Since $1/q(1/k)=p(k)=\int\rmd k\,w(k)$ for each direction, the measure weight in momentum space is
\be\label{wvmulti}
w_\mu(k^\mu)=\left[\frac{p^\mu(k^\mu)}{k^\mu}\right]^2v_\mu\left(\frac{1}{k^\mu}\right)\,.
\ee
In the case of the theory with weighted derivatives, where derivative operators are $\p_\mu\to v_\mu^{-1/2}\p_\mu (v_\mu^{1/2}\,\cdot\,)$, the measure weight $w(k)$ is arbitrary \cite{frc3}. The gravitational and particle-physics actions of these theories can be found in \cite{frc11,frc13}.

The form of the geometric coordinates $q^\mu(x^\mu)$ is dictated by fractal geometry and it is constrained by two requirements: to have an anomalous scaling at small scales (i.e., such that $q$ is not linear in $x$) and to display a discrete scale invariance at possibly even smaller scales \cite{frc2}. In a couple of places below, we will take the example of the isotropic coarse-grained binomial measure
\bs\label{qpst0}
\be\label{qpst00}
q^\mu(x^\mu)\simeq q_*(x^\mu):=x^\mu+{\rm sgn}(x^\mu)\frac{\ell_*}{\a}\left|\frac{x^\mu}{\ell_*}\right|^\a,
\ee
where $0<\a<1$ is a constant and $\ell_*$ is the only characteristic length scale of the measure (more scales correspond to polynomial measures, called multiscale \cite{frc2}). This measure has an anomalous scaling for $|x^\mu|\ll\ell_*$ determined by $\a$ along all spacetime directions (isotropy). Discrete scale invariance has been washed away by a coarse-graining procedure at scales smaller than $\ell_*$ \cite{frc2} and is not apparent in \Eq{qpst00}. In the theory with $q$-derivatives, the conjugate momentum measure reads
\be
p^\mu(k^\mu)\simeq p_*(k^\mu):=\frac{k^\mu}{1+\a^{-1}|\ell_* k^\mu|^{1-\a}}\,.
\ee\es


\subsection{Multifractional from noncommutative}\label{2b}

 We begin the program outlined in the introduction by establishing whether $\kappa$-Minkowski spacetime corresponds to some multifractional spacetime with a certain measure. The symmetry algebra of $\kappa$-Minkowski spacetime is given by the bicross-product $\kappa$-Poincaré algebra and it has been introduced in Refs.\ \cite{majrue,luki} at the beginning of the 1990s. As a first approximation, we can focus on the deformation of the Casimir operator of the $\kappa$-Poincaré algebra: in $D=1+1$ dimensions,
\begin{equation}
\label{kappacas}
\mathcal{C} = -\left(\frac{2}{\lambda}\sinh\frac{\lambda K_0}{2}\right)^2 +e^{\lambda K_0}K^2,
\end{equation} 
where $\la=\lp$ is the Planck length, $K$ and $K_0$ are the generators of, respectively, spatial and time translations in the bicross-product basis and we are restricting to the massless case. Our aim is to find the factorizable measure $\rmd Q_0(X_0)\rmd Q_1(X)$ of position space from the on-shellness relation $e^{-\lambda K_0}\cC=0$ suggested by Eq.\ \eqref{kappacas}. Defining
\begin{equation}
\label{lincoor}
P_0 = \frac{2}{\lambda}e^{-\frac{\lambda K_0}{2}}\sinh\frac{\lambda K_0}{2}\,, \qquad P = K\,,
\end{equation}
we recover the standard relation $-P_0^2+P^2=0$ between the time and the spatial parts of the momentum. We can read off the spacetime coordinates from Eq.\ \eqref{pqmulti}: 
\begin{equation}\label{qq0}
Q = X\,, \qquad Q_0 = \frac{\lambda e^{\frac{\lambda}{2 X_0}}}{2\sinh\frac{\lambda}{2X_0}}=\frac{\la}{1-\rme^{-\la/X_0}}\,.
\end{equation}
Therefore, using the relation \Eq{pqmulti} between conjugate geometric coordinates, we have been able to shift the nontrivial features of the $\kappa$-deformed Casimir \eqref{kappacas} from momentum space to position space. To check that the spacetime dimensionality changes with the scale, we can calculate the Hausdorff dimension $\dh:=\rmd \ln\cV/\rmd\ln R$, where $\cV=\int_{\rm ball}\rmd Q^0\rmd Q^1$ is the volume of a 2-ball of Euclidean radius $\sqrt{X_0^2+X_1^2}=R$. Clearly, the spatial dimension is 1. The Euclideanized time direction is less trivial. Centering the ball at $X_0=0=X$, from Eq.\ \Eq{qq0} one has
\be
\cV\propto \frac{R}{1-\rme^{-\la/R}}\quad \Rightarrow\quad \dh=1-\frac{\la/R}{1-\rme^{\la/R}}\,.
\ee
In $D$ dimensions, one replaces $1\to D-1$. In the infrared (IR, $|\la/R|\ll 1$, large scales and long time intervals), we get standard spacetime with $Q_0\simeq X_0$, $\cV\sim R^D$ and $\dh\simeq D-1+1=D$. In the ultraviolet (UV, $|\la/R|\gg 1$, small scales and short time intervals), the time direction becomes degenerate, $Q_0\simeq \la(1+\rme^{-\la/X_0})\simeq \la$, and $\dh\simeq D-1+0=D-1$. Thus, the Hausdorff dimension runs from $D-1$ to $D$ monotonically. In 4 dimensions, it runs from $3$ to $4$.

Another useful geometric indicator is the spectral dimension of spacetime (see, e.g., \cite{CMNa} for an introduction), defined as $\ds:=-2\rmd\ln\cP(\s)/\rmd\ln\s$, where $\s$ is a $({\rm length})^2$ parameter representing the probed scale and, in the multifractional theory with $q$-derivatives, $\cP(\s)=\int\rmd^DP\,\exp[-Q^0(\s)\,P_\mu P^\mu]\propto [Q^0(\s)]^{-D/2}$ \cite{frc7}. Then, $\ds=D\la/[(\rme^{\la/\s}-1)\s]$. In the IR ($\la/\s\ll 1$), $\ds\simeq D$, while in the UV ($\la/\s\gg 1$) $\ds\simeq 0$.

However, the multifractional spacetime found from the Casimir operator is not $\kappa$-Minkowski spacetime. An easy way to see this is to compare the measure in momentum space, which is different: factorizable in the multifractional case (in order to have an invertible Fourier transform \cite{frc3,frc11}) and nonfactorizable in the noncommutative case. Also, the running of $\ds$ found above is not the dimensional flow of $\kappa$-Minkowski spacetime where, for the bicross-product Casimir, the spectral dimension \emph{decreases} from the UV to the IR \cite{ArTr1}. Therefore, the Casimir alone cannot establish a duality between $\kappa$-Minkowski and a multifractional spacetime, although it does correspond to the dispersion relation of a multiscale spacetime. This spacetime is not multifractal because the measure $Q^0(X^0)$ in Eq.\ \Eq{qq0} does not correspond to a fractal geometry \cite{frc2}. The same conclusion is reached after computing the walk dimension and noting that it does not combine with the Hausdorff and spectral dimension in the way it should for fractals \cite{trtls}.


\subsection{Noncommutative from multifractional}\label{2c}

The factorizable measure of multifractional models is the main obstacle towards establishing a duality between them and noncommutative spacetimes. However, commutative multiscale theories with nonfactorizable measures \cite{fra1,fra2,fra3} were shown to be not very manageable in early studies of fractal spacetimes on a continuum \cite{revmu}, which was the reason to propose the factorizable measures of modern multifractional theories \cite{frc2,frc1}. Since the technical problems entailed in multiscale nonfactorizable measures seem unavoidable, and since $\kappa$-Minkowski \emph{is} a multiscale theory (by definition) where nonfactorizability issues are solved with the elegant machinery of noncommutative products, we might as well regard {noncommutative spacetimes as the natural generalization of multifractional spacetimes to nonfactorizable measures}. In this case, both classes of theories are multiscale but the landscape of noncommutative models might contain the landscape of multifractional spacetimes. If this conjecture were true, one should be able to write a nontrivial phase-space Heisenberg algebra for any of the four known multifractional theories. The theory with ordinary derivative does not have a well-defined momentum transform and has therefore been regarded as a multiscale toy model; we do not expect it to correspond to any noncommutative spacetime. The theory with fractional derivative is still under construction and we cannot say much about its relation with noncommutative models. The theory with $q$-derivatives and that with weighted derivatives are the best studied and we can work directly on them. The following calculation proves the conjecture ``noncommutative implies multifractional'' \emph{wrong}. In other words, despite some remarkable similarities at the level of the spacetime measure, noncommutative and multifractional models constitute separate, nonoverlapping regions in the landscape of multiscale theories.

Consider the multifractional theory with $q$-derivatives. If it corresponded also to a noncommutative spacetime, then we should be able to derive the Moyal product from the product of functions of the geometric coordinates $q^\mu(x^\mu)$ defining the theory. The opportunity of finding the $\star$-product in this way resides in the nonlinearities brought by both the coordinates $q^\mu(x^\mu)$ and their conjugate momenta $p_\mu(k_\mu)$. Thus, let us consider the composition of two plane waves
\begin{equation}\label{plwav}
e^{ip_\mu(k_\mu)\,q^\mu(x^\mu)}\,e^{i p_\nu(\widetilde{k}_\nu)\,q^\nu(x^\nu)},
\end{equation}
where index contraction follows the Einstein convention ($p_\mu q^\mu=\eta_{\mu\nu}p^\mu q^\nu$) and, for our purposes, the coordinate profiles are given by Eq.\ \Eq{qpst0}. Although the full multifractional profiles are more complicated, the binomial example is enough. Momenta $p_\mu(k_\mu)$ and coordinates $q^\mu(x^\mu)$ are nonlinear functions of $k_\mu$ and $x^\mu$, respectively. Let us suppose, for simplicity, that the measure is deformed only in the spatial part, i.e., $q^0 \equiv x^0$ and $p_0 \equiv k_0$. Our aim is to interpret Eq.\ \eqref{plwav} as the Moyal product $e^{ik_\mu x^\mu}\star e^{i\widetilde{k}_\nu x^\nu}$ of two plane waves on $x$-space. Plugging Eq.\ \Eq{qpst0} into Eq.\ \eqref{plwav},  expanding for small momenta, and taking the resulting expression as our definition of the $\star$-product, in $1+1$ dimensions we get
\ba
e^{ik_\mu x^\mu}\star e^{i\widetilde{k}_\nu x^\nu} &:=& \exp\left[i(k_\mu+\widetilde{k}_\mu)x^\mu +i\frac{\ell_*}{\alpha}(k_1+\widetilde{k}_1)\left|\frac{x_1}{\ell_*}\right|^\alpha\right.\nonumber\\
&&\left.-i\left(\frac{k_1}{|\ell_* k_1|^{\alpha-1}}+\frac{\widetilde{k}_1}{|\ell_* \widetilde{k}_1|^{\alpha-1}}\right)\frac{x_1}{\alpha}\right]\!.\label{starproddef}
\ea
The final step consists in using the above definition to find the corresponding noncommutative theory. This can be done by means of a Weyl map, which is an isomorphism between a given noncommutative algebra for the spacetime coordinates $X^\mu$ and a corresponding $\star$-product (or Moyal product). In other words, a Weyl map $\Omega$ is a one-to-one correspondence between a noncommutative theory and a commutative theory with a nontrivial multiplication rule. This means that, using a Weyl map $\Omega$, we can write the product of two functions $F(X^\mu)$ and $G(X^\nu)$ depending on noncommutative coordinates $X^\mu$ in terms of a nontrivial multiplication rule between two functions $f(x^\mu)$ and $g(x^\nu)$ of the commutative coordinates, i.e., $F(X^\mu) G(X^\nu) = \Omega[f(x^\mu)\star g(x^\nu)]$. We hereby introduce a suitable Weyl map defined by
\ba
e^{ik_\mu x^\mu}\star e^{i\widetilde{k}_\nu x^\nu} &:=& \Omega^{-1}\left(e^{ik_\mu X^\mu}e^{i\widetilde{k}_\nu X^\nu}\right)\nonumber \\
&\simeq& \Omega^{-1}\left(e^{i(k_\mu+\widetilde{k}_\mu)X^\mu-\frac{k_\mu \widetilde{k}_\nu}{2}[X^\mu,X^\nu]}\right)\nonumber \\
&=& \Omega^{-1}\left(e^{i(k_\mu+\widetilde{k}_\mu)X^\mu+\frac{k_0 \widetilde{k}_1 - k_1\widetilde{k}_0 }{2}[X^1,X^0]}\right),\nonumber\\\label{temp1}
\ea
where we have used the first-order approximation of the Baker--Campbell--Hausdorff (BCH) formula. Equating this with Eq.\ \eqref{starproddef}, we finally obtain the commutation rule
\ba
[X^1,X^0] &=& \frac{2i}{k_0 \widetilde{k}_1 - k_1\widetilde{k}_0}\left[\frac{\ell_*}{\alpha}(k_1+\widetilde{k}_1)\left|\frac{X^1}{\ell_*}\right|^\alpha\right.\nonumber\\
&&\left.-\left(\frac{k_1}{|\ell_* k_1|^{\alpha-1}}+\frac{\widetilde{k}_1}{|\ell_* \widetilde{k}_1|^{\alpha-1}}\right)\frac{X^1}{\alpha}\right].\label{xxbad}
\ea
If we wrote, for instance, the noncommutative Lagrangian of a scalar field with this result, then by construction we would obtain the scalar-field Lagrangian of the $q$-theory approximately.

However, Eq.\ \Eq{xxbad} is ill-defined because it depends on the momenta of both plane waves, while it should be momentum independent. The explicit reference to plane waves' momenta prevents us from interpreting Eq.~\eqref{xxbad} as a general noncommutative spacetime algebra that should hold for any number of waves. This happens because we imposed the commutator to give the nonlinear terms coming from the BCH formula. For a well-defined noncommutative theory there is a mutual compatibility between the $\star$-product, the Weyl map $\Omega$ and the noncommutativity of $X^\mu$. In particular, the $\star$-product matches the nonlinear functions of the momenta appearing in the terms of the BCH expansion [see the last line of Eq.\ \eqref{temp1}] in such a way that the commutator involving $X^\mu$ does not depend on momenta. Clearly, it does not happen in the case we are analysing here. Moreover, both \Eq{starproddef} and \Eq{xxbad} are completely \emph{ad hoc} formul\ae\ constructed for the composition of two plane waves and they would not work for three or more phases. All these problems stem from the factorizability of the measure of the $q$-theory. There is, in fact, a clear tension between Eqs.\ \Eq{starproddef} and \Eq{temp1}: while the first is a factorized composition of position and momentum coordinates, the second tends to mix the momenta of both waves. Forcing the definition \Eq{starproddef} results in the expression \Eq{xxbad}.

That the form of the multifractional measure is the main problem for an interpretation of multifractional theories as noncommutative ones can be seen in another way. Consider the scalar-field action in the $q$-theory in $1+1$ dimensions:
\ba
S_q &=& -\frac12\int d^2q\left(\p_{q^\mu}\phi\p^{q^\mu}\phi+m^2\phi^2+\frac{2\sigma}{n!}\phi^n\right)\nonumber\\
    &=& \frac12\int dq^0 dq^1\left[(\p_{q^0}\phi)^2-(\p_{q^1}\phi)^2-m^2\phi^2-\frac{2\sigma}{n!}\phi^n\right]\nonumber\\
		&=&\frac12\int dx^0 dx^1\left[\frac{v_1}{v_0}(\p_0\phi)^2-\frac{v_0}{v_1}(\p_1\phi)^2-v_0v_1m^2\phi^2\right.\nonumber\\
		&&\qquad\qquad\left.-v_0v_1\frac{2\sigma}{n!}\phi^n\right]\label{scac}
\ea
and let us compare it with the scalar-field action in a generic (i.e. without specifying any specific form for the $\star$-product) noncommutative theory: 
\be
S_\star = -\frac12\int d^2x\left(\p_{\mu}\phi \star \p^{\mu}\phi+m^2\phi \star \phi+\frac{2\sigma}{n!}\phi\star\phi\star ... \star \phi\right).\label{scacnoncomm}
\ee
In the action $S_q$, we have done easy manipulations in order to shift the nontrivial form of the $q$-measure as well as of the $q$-derivatives to prefactors in front of the fields. In this way, since in a noncommutative theory the $\star$-product between fields produce nontrivial prefactors, we can check whether it is possible to match deformations in $S_q$ with those carried by the $\star$-products in $S_\star$. However, this is not the case. In $S_q$, there are three terms quadratic in the field $\phi$ but all of them have different measure prefactors given by the combinations of the profiles $v_0(x^0)$ and $v_1(x^1)$. In $D$ dimensions, the $\mu$th component of the kinetic term has a ``deformation'' $v_0 v_1\cdots(1/v_\mu)\cdots v_{D-1}$, while the mass term has a $v_0\cdots v_{D-1}$ prefactor. It is then difficult to read a $\star$-product in this type of action, since terms in $S_\star$ with the same number of fields (e.g. kinetic and mass term) have the same deformation because they are all of the form $\phi \star \phi$ and the derivatives of the kinetic term do not affect the $\star$-product. This is a general feature of noncommutative theories that does not fit the structure of multifractional actions. 

The same conclusion can be reached in all other multifractional theories with factorizable measures. For instance, in the theory with weighted derivatives the free scalar-field case is trivial because, after a field redefinition $\phi\to \phi/\sqrt{v_0v_1}$, the $O(\phi^2)$ part coincides with a commutative theory (see Ref.\ \cite{frc6} for the details of the dynamics in $D$ dimensions). This is not an issue \emph{per se} because one could invoke the trace property on the free part and concentrate on nonlinear field terms. The interaction $\phi^n$ has exactly the same structure as in Eq.\ \Eq{scac} and its deformation $v_0v_1$ could be used as a $\star$-product, were it not for the fact that interacting noncommutative field theories are not easy to work out. Although we do not try this calculation here, we do not foresee any way to avoid the factorizability problem.


\section{Noncommutative and multifractional}\label{3}

Although we cannot interpret multifractional spacetimes as noncommutative, we can make them so and study the corresponding deformed symmetry algebras. Instead of a direct construction, we follow a more attractive path which, in generic terms, starts from a noncommutative symmetry algebra and leads to a multifractional measure. We begin with a special case and then move to the general one.


\subsection{Multifractional spacetimes from \texorpdfstring{$\kappa$}{}-Minkowski phase-space algebra}\label{3a}

Working in $D=1+1$ dimensions, we can denote with $(Q, Q_0, P, P_0)$ the phase-space operators of the multifractional theory with $q$-derivatives with a generic nontrivial weight measure given by $dQ_0 dQ = dX_0 dX v(X)$. We assume that such a deformed measure only depends on the spatial coordinate $X$, while the time part is left unmodified (i.e., it has a trivial weight). This assumption is dictated only by the aim of the following calculation, which is to reproduce the $\kappa$-Minkowski algebra. Of course, one can conceive the general case with a nontrivial time measure and repeat the procedure detailed below. In that case, one will find a more general noncommutative spacetime that collapses to $\kappa$-Minkowski in the limit $Q_0(X_0)\to X_0$. The calculation would be complicated by the presence of commutators $[f_1(X_0),f_2(X)]$ between functions of operators, which can be written as infinite series once $f_{1,2}$ are known \cite{TrVa}.

By definition, the geometric coordinates obey the Heisenberg algebra
\begin{equation}
\label{mltps}
[Q,P] = i, \quad [Q_0,P_0] = -i, \quad [Q,P_0] = [Q_0,P] = 0,
\end{equation}
and they are related to the phase space generated by $(X, X_0, K, K_0)$ in the following way:
\begin{equation}\label{relkml}
Q = \int \rmd X\,v(X), \quad Q_0 = X_0, \quad P = \frac{1}{v(X)}K, \quad P_0 = K_0\,,
\end{equation}
where $v$ is the measure weight in the spatial direction. The third expression is a consequence of imposing the canonical commutation relations $[Q,P]=\rmi$ and $[X,K]=\rmi$, which are the quantum counterpart of the classical canonical relation \Eq{pqmulti}.

We want to prove that the multifractional weight is given by $v(X) \propto |X|^{-1}$ if $X$ and $X_0$ are $\kappa$-Minkowski coordinates, i.e.,
\be\label{xx0mink}
[X,X_0] = \rmi \lambda X\,.
\ee
Such a result, that establishes a connection between multifractional and noncommutative spacetimes, was first derived in Ref.\ \cite{ACOS}. However, in that case the analysis was done in position space and by using the $\star$-product to find a map between the set of $(Q,Q_0)$ coordinates and $(X, X_0)$. Information on the multifractional momentum space was not used and this permitted to keep the multifractional side of the correspondence arbitrary. On the other hand, here we find the same outcome in a more compact way just using the deformed Heisenberg algebra of the $\kappa$-Minkowski phase space, but specifying the multifractional theory to be the one with $q$-derivatives. 

The $\kappa$-Heisenberg algebra is given by the commutation relations \cite{kappaps} 
\begin{eqnarray} \label{kps}
&&[X,K] = i, \quad [X_0,K_0] = -i,\\ &&[X,K_0] = 0, \quad  [X_0,K] = i\lambda K,
\end{eqnarray}
as one can easily check by computing the Jacobi identities involving the phase-space operators and taking into account \Eq{xx0mink}.

The explicit form of the measure weight $v(X)$ can be derived thanks to the two sets of commutators \eqref{mltps} and \eqref{kps}. To this aim, let us consider the commutation relation between time $Q_0$ and the spatial momentum operator $P$:
\ba
0 &=& [P, Q_0] = [\frac{1}{v(X)}K, X_0]\nonumber\\
&=&\frac{1}{v(X)}[K, X_0] + [\frac{1}{v(X)},X_0]K\nonumber\\
&=& \frac{1}{v(X)} (-\rmi\lambda K) -\frac{v'(X)}{v^2 (X)}[X, X_0] K\nonumber\\
&=&-\frac{\rmi\lambda}{v(X)}\left[1+\frac{v'(X)}{v(X)}X\right] K\,,
\ea
where $v'(X) =dv(X)/dX$ and we have used the third expression in Eq.\ \eqref{relkml} and the phase-space commutators \Eq{kps}. Notice that the ordering between $X$ and $K$ is nontrivial because they are noncommuting variables. Integrating over $X$ and introducing a length scale $\la$ to keep $v$ dimensionless, we get
\begin{equation}\label{multmeas}
-\int\frac{dX}{X} = \int \frac{dv}{v} \quad\Rightarrow\quad v(X) = \frac{\la}{|X|}\,,
\end{equation}
which is exactly the measure found in Ref.\ \cite{ACOS}. Apart from the shortness of this novel derivation, the main advantage comes from the fact that we have not assumed any specific form for the integration measure on $\kappa$-Minkowski spacetime, contrary to the analysis of Ref.\ \cite{ACOS}. There, the argument was based on a comparison of the fractional measure with the $\kappa$-Minkowski cyclic-invariant measure, which has the drawback of breaking the relativistic symmetries (see, e.g., \cite{cyckappameas}). Here we have found the measure \eqref{multmeas} relying only on the commutators of the phase space of both multiscale \eqref{mltps} and $\kappa$-Minkowski \eqref{kps} variables. In this way, we have not been forced to introduce a symmetry-breaking measure on $\kappa$-Minkowski spacetime.

The measure weight $v(x)\sim 1/|x|$ arises as the ultraviolet limit of a multifractional measure with logarithmic oscillations. In this limit, the fundamental scale $\ell_\infty$ appearing in the oscillatory part is factored out of the asymptotic measure as an overall constant. Thus, the theoretical problem of the disappearance of the Planck length in the $\kappa$-Minkowski cyclic-invariant measure was solved in \cite{ACOS} by regarding $\kappa$-Minkowski spacetime as the limit of noncommutative multifractional Minkowski spacetime and by identifying $\ell_\infty$ with the Planck scale. This embedding would be fully valid only if the symmetries of $\kappa$-Minkowski exactly matched those of the multifractional $q$-theory. Here we checked this correspondence at the level of the Heisenberg algebra and, in the next subsection, we will give another proof at the level of the Poincar\'e algebra. Therefore, the geometrical and physical interpretation of \cite{ACOS} is confirmed. Note that there is no contradiction between this result and the fact that we cannot identify multifractional field theories with noncommutative field theories, first because the embedding of $\kappa$-Minkowski in the multifractional framework is at the level of spacetime, not of field theory; and, second, because such embedding is of a noncommutative spacetime within another, while the negative results of the previous section involve noncommutative theories on one hand and commutative multifractional theories on the other hand.


\subsection{Noncommutative spacetimes from multiscale deformed symmetries}\label{3b}

In this subsection, we start from the multifractional $q$-theory and recast it as a noncommutative spacetime with exactly the same symmetries. By definition, the dynamics of this theory in the absence of curvature is invariant under the so-called $q$-Poincaré symmetries
\begin{equation}\label{qpoi}
q^\mu(x'^\mu) = \Lambda^\mu_{\ \nu} q^\nu(x^\nu) + a^\mu\,,
\end{equation}
which correspond to highly nonlinear transformations of the $x$-coordinates. This means that, in the $q$ position space, we have the undeformed Poincaré commutators between the classical generators $\cN$ and momenta $(P_0,P)$ of, respectively, infinitesimal boosts and time-space translations:
\begin{equation}
\label{qPoinc}
[\mathcal{N},P] = iP_0\,, \qquad [\mathcal{N},P_0] = iP\,, \qquad [P_0,P] = 0\,,
\end{equation}
where
\be\nonumber
\mathcal{N}= i\left(Q\frac{\partial}{\partial Q_0}-Q_0\frac{\partial}{\partial Q}\right)\,,\quad P_0=i\frac{\partial}{\partial Q_0},\quad P=-i\frac{\partial}{\partial Q}\,.
\ee
On the other hand, these $q$-Poincaré commutators generate the nonlinear transformations \Eq{qpoi} on the $X$ position space. In order to make this manifest, we derive the symmetry algebra expressed in terms of the momenta $(K_0,K)$. To this end, we consider the simplified case in which only the spatial part of the measure is modified. Then, $P_0 = K_0$ and $P=P(K)$ is determined by the geometric coordinates in position space via Eq.\ \Eq{pqmulti}. In terms of the momenta $(K_0,K)$, the symmetry algebra is
\be\label{nonlinPa}
[\mathcal{N},K] = \frac{iK_0}{w(K)},\qquad [\mathcal{N},K_0] = \rmi P(K), \qquad [K,K_0] = 0,
\ee
where, according to Eq.\ \Eq{wvmulti}, $w(K)=(P^2/K^2) v(1/K)$. These commutation relations reduce to the usual Poincaré algebra if we send to infinity the deformation parameter appearing in $w\to 1$ and $P\to K$. For instance, for the operatorial version of the binomial measure \Eq{qpst0} 
\bs\label{qpst}\ba
Q(X)&=& X+{\rm sgn}(X)\frac{\ell_*}{\a}\left|\frac{X}{\ell_*}\right|^\a,\\
P(K)&=& \frac{K}{1+\a^{-1}|\ell_* K|^{1-\a}}\,,
\ea\es
one has
\be\label{vwexpl}
v(X)=1+\left|\frac{X}{\ell_*}\right|^{\a-1},\quad w(K)=\frac{1+|\ell_* K|^{1-\a}}{(1+\a^{-1}|\ell_* K|^{1-\a})^2},
\ee
and the limit giving the standard Poincaré algebra is $|\ell_*/X| \rightarrow 0 \leftarrow |\ell_*K|$ (vanishing fundamental length scale at which multiscale effects become apparent).

Interestingly, the deformation we have obtained is given by nonlinear functions of the generators of translations (i.e., $K$ and $K_0$) on the $X$ position space. These kinds of modifications are those studied to characterize the relativistic symmetries of noncommutative spacetimes (see Ref.\ \cite{defpa} for a recent review on generalized deformations of the Poincaré algebra in the framework of quantum groups). In the light of this analogy, we want to determine what type of noncommutativity of the coordinates $(X_0,X)$ is implied by \eqref{nonlinPa}. Our strategy is to derive the commutation relations involving the set of operators $(\mathcal{N},K_0,K,X_0,X)$ from the known commutators of both the $q$-Poincaré algebra \eqref{qPoinc} and the $Q$ phase space. Then, we will look for the outcome of the commutator $[X,X_0]$ needed to satisfy all the Jacobi identities.

Let us start by deriving the commutators between the boost operator $\mathcal{N}$ and $(X_0,X)$. They can be obtained from the corresponding commutators on the $Q$ space, which are by definition
\begin{equation}
\label{boostcoord}
[\mathcal{N},Q_0] = iQ, \qquad [\mathcal{N},Q] = iQ_0\,,
\end{equation}
giving the desired commutation relations $[\mathcal{N},X_0]$ and $[\mathcal{N},X]$:
\begin{equation}\label{tempo0}
[\mathcal{N},X_0] = \rmi Q(X), \qquad [\mathcal{N},X] = \rmi X_0 v^{-1}(X)\,.
\end{equation}
Given the above deformed actions of $\mathcal{N}$ on the coordinates, one can now derive the commutator between spacetime coordinates by requiring the validity of the Jacobi identity involving $(\mathcal{N},X,X_0)$: 
\ba
0 &=& [[\mathcal{N},X],X_0]+[[X_0,\mathcal{N}],X]+[[X,X_0],\mathcal{N}]\nonumber\\
  &=& \rmi X_0[v^{-1}(X),X_0]+[[X,X_0],\mathcal{N}]\nonumber\\
	&=& -\rmi X_0[X,X_0]\frac{v'}{v^2}+[[X,X_0],\mathcal{N}]\,.\label{tempo}
\ea
At this point, we make two mutually exclusive \emph{Ansätze}: either
\be
[X,X_0]=\rmi h(X_0)
\ee
or 
\be\label{xx0f}
[X,X_0]=\rmi f(X)\,.
\ee
In the first case, Eq.\ \Eq{tempo} and the first commutator in \Eq{tempo0} give $X_0 h(X_0)v'(X)/v^2(X)=Q(X)h'(X_0)$, which is solved by
\be\nonumber
h(X_0)=\b\rme^{X_0^2/(2l^2)}\,,\qquad Q(X)=\sqrt{2}l\,{\rm Erf}^{-1}\left(\sqrt{\frac{2}{\pi}}\frac{X}{l}\right),
\ee
where $\b$ is a dimensionless constant, $l$ is a constant length and ${\rm Erf}^{-1}$ is the inverse error function. This noncommutative spacetime is compact and has a very strange behaviour: it has a canonical position-space algebra in the double early-time limit $|X_0/l|\ll 1$ and UV limit $|X/l|\ll 1$ (where $Q\simeq X$). Since it does not possess a well-defined IR limit, we discard this solution.

Case \Eq{xx0f} is more appealing. From Eq.\ \Eq{tempo} and the second commutator in \Eq{tempo0}, we have $v'/v=-f'/f$, hence $f=\la^2/v$, where $\la$ is a constant length:
\begin{equation}\label{noncommx}
[X,X_0] = \frac{\rmi\la^2}{v(X)}\,.
\end{equation}
Fortunately, the measure weight $v(X)$ is unconstrained and it can take the standard form in multifractal spacetimes with $q$-derivatives [in the absence of log oscillations, Eq.\ \Eq{vwexpl}]. If $\la=0$, the algebra of the coordinates is trivial, $[X,X_0]=0$ and position space is commutative. If $\la\neq 0$, then $Q$ position space is canonical. In fact, from the definition of geometric coordinates it follows directly that
\be\label{canoqq}
[Q,Q_0] = i\la^2\,.
\ee
The nature of position space depends on whether one imposes $\la=0$ (commutativity) or $\la\neq 0$ (noncommutativity). Note that for $v(X)=\la/X$, Eq.\ \Eq{noncommx} reproduces the $\kappa$-Minkowski algebra \Eq{xx0mink}. Thus, up to an absolute value we have obtained the same result of the previous subsection, but using the Poincaré algebra instead of the Heisenberg one. Repeating the procedure we adopted to derive Eq.\ \eqref{nonlinPa} and considering the Jacobi identity for $\cN$, $X_0$ and $K$, the remaining commutators read
\bs\ba
&&[K,X] = -\frac{\rmi}{v(X)\,w(K)},\qquad [K_0,X_0] = i,\\
&&[K_0,X] = 0, \qquad [K,X_0] = 0\,.
\ea\es
Equipped with these commutators, one can finally check that all the Jacobi identities are satisfied.

The choice $\la\neq 0$ in Eq.\ \Eq{noncommx} defines a noncommutative extension of the multifractional theory under examination. In order to complete this extension, we need to identify a suitable Weyl map. After having found a correspondence between the noncommutativity given by Eq.\ \eqref{noncommx} on the $x$-space and the canonical noncommutative $q$-space, it is immediate to write down the $\star_q$-product for a canonical spacetime with \Eq{canoqq}:
\ba
f_p(q^0,q)\star_q g_k(q^0,q) &=& \Omega^{-1}_{q}[f_p(Q^0,Q)g_k(Q^0,Q)]\nonumber\\
&=& e^{i(p_\mu + k_\mu)q^\mu} e^{-i\la^2 p^0k},\label{prodq}
\ea
where $\mu = 0,1$. Such a Weyl map allows us to work with functions depending on commutative coordinates $(q_0,q)$ equipped with the $\star_q$-product \eqref{prodq}. For instance, the action for a real scalar field $\phi$ with self-interaction reads
\ba
S_q^\star &=& -\int dq^0dq\left(\frac{1}{2}\partial_{q^\mu}\phi \star_q \partial^{q^\mu}\phi+\frac{m^2}{2}\phi\star_q\phi\right.\nonumber\\
&&\left.\qquad\qquad+\frac{\sigma}{n!}\phi\star_q\cdots\star_q\phi\vphantom{\frac{\partial_{q^\mu}\phi \star_q \partial^{q^\mu}\phi}{2}}\right).
\ea
The same line of reasoning applies also to the $x$ position space but with more technicalities due to the form of Eq.\ \eqref{noncommx}. By definition of the Weyl map, we know that $f_p(X^0,X)g_k(X^0,X) = \Omega_x[f_p(x^0,x)\star_x g_k(x^0,x)]$, where the coordinates $(x^0,x)$ are commutative while $(X^0,X)$ obey Eq.\ \eqref{noncommx}. Then, we can express functions of noncommutative coordinates as inverse Fourier transforms of commuting functions on momentum space, i.e., $f(X^0,X) = (2\pi)^{-1}\int dp^0dp\,e^{ip_\mu x^\mu} \bar{f}(p^0,p)$. Thus, in order to find the $\star_x$-product explicitly, we must be able to compute the product of phases such as $e^{ip_\mu X^\mu}e^{ik_\nu X^\nu}$ depending on noncommuting operators. This can be done by exploiting the BCH lemma that, in general, gives such a product in terms of the sum of the two operators plus an infinite series of corrections. The latter are combinations of the commutators between the operators: $\exp(ip_\mu X^\mu)\exp(ik_\nu X^\nu)= \exp[i(k_\mu+p_\mu)X^\mu-k_\mu p_\nu[X^\mu,X^\nu]/2+O(\la^4)]=\exp\{i(k_\mu+p_\mu)X^\mu+i\la^2(k^0p-kp^0)/[2v(X)]+O(\la^4)\}$, where we used Eq.\ \eqref{noncommx} and we restricted only to the first-order correction term. Unfortunately, in the case of Eq.\ \eqref{noncommx} we do not have a simplified version of the BCH formula. This prevents us from finding explicitly the $\star_x$-product at all orders in $\la$ which, thus, can be introduced only in a formal way (i.e., order by order). 


\section{Multiscale hypersurface deformation algebra}\label{4}

So far, we have ignored gravity and considered spacetimes embedded in flat Minkowski. Turning gravity on, we can extract interesting information about spacetime symmetries in the curved case. 

In the Hamiltonian formulation of general relativity, the general covariance of the theory is encoded in the algebra closed by the scalar ($H[N]$) and vector ($D[N^{i}]$) constraints, the so-called hypersurface-deformation algebra (HDA) \cite{ADM}:
\ba
&&\{D[M^{k}],D[N^{j}]\}=D[\mathcal{L}_{\vec{M}}N^{k}],\nonumber\\
&&\{D[N^{k}],H[M]\}=H[\mathcal{L}_{\vec{N}}M],\label{hda}\\
&&\{H[N],H[M]\}=D[h^{jk}(N\partial_{j} M-M\partial_{j} N)],\nonumber
\ea
where $H[N]$ and $D[N^{i}]$ depend, respectively, on the lapse function $N$ and the shift function $N^{i}$ and $\cL$ is the Lie derivative.

Multiscale spacetimes depart from classical Riemannian geometry due to the introduction of a nontrivial integro-differential structure independent of the metric structure. A natural question, which we answer here for the first time, is whether the HDA should be deformed in this framework. Moreover, recently there has been a growing effort in studying quantum deformations of Eqs.\ \eqref{hda} in the context of effective models motivated by loop quantum gravity \cite{bojopaily,BBCGK}. Therefore, it is interesting to compare the LQG modifications in the effective-dynamics approach with possible modifications of the HDA in the multifractional approach. 

In the previous sections, we have seen that multifractional measures in the Minkowski embedding produce nonlinear deformations of the Poincaré algebra. In general, the Poincaré algebra can be obtained as the flat-spacetime limit of the HDA, a fact that crucially helped to find an agreement between $\kappa$-Poincaré and the HDA with LQG holonomy corrections (see Refs.\ \cite{bojopaily1,hdaPa}). This suggests to look for a possible connection between LQG in the effective-dynamics approach and multifractional theories. In this section, we derive the HDA in two different multifractional models: the theory with $q$-derivatives and that with weighted derivatives.


\subsection{Theory with \texorpdfstring{$q$}{}-derivatives}

Gravity in multifractional theories has been studied in Ref.\ \cite{frc11}. The case of the $q$-theory is simple and amounts to replacing $x^\mu \rightarrow q^\mu(x^\mu)$ everywhere in the standard Einstein--Hilbert action of general relativity. Despite its simplicity, this replacement gives rise to a nontrivial physics because it introduces a preferred frame where all observables should be computed \cite{trtls,frc11}. It is easy to guess that the constraint algebra has the same form of Eq.\ \eqref{hda}, with the difference that coordinates now are the composite objects $q^\mu(x^\mu)$. However, as a consequence, neither the first-class constraints \eqref{hda} nor the Lie derivatives therein are the standard ones. Since the spatial $q$-derivatives can be expressed as $\p_{q_i} = v_i^{-1}(x^i)\p_i$ (where $\partial_i=\p/\p x^i$), we can write explicitly the $q$-HDA as 
\ba
&&\{D^q[M^{k}],D^q[N^{j}]\}=D^q\left[\frac{1}{v_j(x^j)}(M^j\partial_j N^{k}-N^j \partial_j M^k)\right],\nonumber\\
&&\{D^q[N^{k}],H^q[M]\}=H^q\left[\frac{1}{v_j(x^j)} N^j\partial_j M\right],\label{qhda}\\
&&\{H^q[N],H^q[M]\}=D^q\left[\frac{h^{jk}}{v_j(x^j)}(N\partial_{j} M-M\partial_{j} N)\right],\nonumber
\ea
where the index of the deformed measure weight $v_j$ is inert and it is not contracted with other indices. We stress that the constraints $H^q[N]$ and $D^q[N^{k}]$ generate time translations and spatial diffeomorphisms of the geometric coordinates $q^\mu(x^\mu)$, which means that these are not the usual time translation and diffeomorphisms, as it would become evident when turning to $x$-spacetime. 

Thus, all Poisson brackets acquire the same anisotropic deformation in the right-hand side. Such a result is not compatible with the LQG modifications of the HDA in the effective-dynamics approach because, in the latter case, spatial diffeomorphisms are unmodified (i.e., both $\{D,H\}$ and $\{D,D\}$ remain untouched). On the other hand, the scalar part $\{H^q,H^q\}$ of Eq.\ \eqref{qhda} can be compared with the analogous LQG bracket
\be\label{betde}
\{H[N],H[M]\}=D\left[\b h^{jk}(N\partial_{j} M-M\partial_{j} N)\right],
\ee
where $\b$ is a phase-space background-dependent function. Although one might naively identify the LQG deformation function $\beta = 1/v_i(x^i)$ with the inverse of the multifractional spatial measure weight, we also have deformations in the other brackets. Another point of departure comes from the fact that the $q$-deformation \Eq{qhda} of the HDA is background independent: it consists only in the measure of the anomalous geometry, which is completely independent of the metric structure. Finally, while $\b$ can change sign in different regimes (an effect often interpreted as a spacetime signature change), $1/v$ is always positive definite.

A deformed HDA and the related signature-change effect appear only when cancellation of quantum anomalies is imposed in the LQG algebra. In two cosmological approaches to loop quantum gravity, based on a dressed metric \cite{AAN1,AAN2,AAN3,AgMo} or on a hybrid quantization scheme \cite{CFMO,CGMM,deBO,CGMM2}, no such deformation is found ($\b=1$ in \Eq{betde}). The multifractional theory with $q$-derivatives also differs from these cases, since all $q$-Poisson brackets are deformed and the gravitational physics is qualitatively different from the LQG one \cite{frc11}. We conclude that, regardless of the quantization scheme adopted, the HDA of loop quantum gravity (and, in particular, loop quantum cosmology) and of the multifractional $q$-theory are physically inequivalent.


\subsection{Theory with weighted derivatives}

In the multifractional theory with weighted derivatives, the gravitational field behaves quite differently. After a frame choice, a conformal transformation of the metric and some field redefinitions, it is possible to write the gravitational action of the system as the standard Einstein--Hilbert action plus a rank-0 function $\phi(x)$ that looks like a scalar field \cite{frc11}. Since the form of the HDA is insensitive to the specific matter content of the theory, one might think that the gravitational and the scalar parts should satisfy separately the classical HDA \eqref{hda}. However, $\phi=\phi[v^\mu(x^\mu)]$ is \emph{not} a scalar field, since it is a nondynamical function of the measure. 

The super-Hamiltonian constraint can be written as $H[N] = H_0[N] + H_\phi[N]=\int d^3x\,N(\cH_0+\sqrt{h}\mathcal{H}_\phi)$, where $h$ is the determinant of the spatial metric,
\be
\cH_0=\frac{\pi_{lk}\pi^{lk}}{\sqrt{h}}-\frac{\pi^{2}}{2\sqrt{h}}-{}^{(3)}R\sqrt{h}
\ee
is only metric dependent and the density $\mathcal{H}_\phi$ is both metric and measure dependent. The diffeomorphism constraint is the usual one, $D[N^{k}]=-2\int d^{3}x\,N^{k}h_{kj}D_{l} \pi^{lj}$. Since there are no dynamical degrees of freedom associated with $\phi$, there is no conjugate momentum $\pi_\phi$. Thus, when computing the Poisson brackets \eqref{hda}, the only contribution of the measure-dependent $\phi$ part is given by the last two pieces in
\ba
\{H[N],H[M]\}&=& \{H_0[N],H_0[M]\}\nonumber\\
&&+\int d^3x\,N(x)\int d^3y\,M(y)\nonumber\\
&&\times\{\mathcal{H}_0(x),\sqrt{h}\}\mathcal{H}_\phi (y)\nonumber\\
&&+\int d^3x\,N(x)\int d^3y\,M(y)\nonumber\\
&&\times\mathcal{H}_\phi (x)\{\sqrt{h},\mathcal{H}_0(y)\}\,.
\ea
However, it is easy to realize that the last two Poisson brackets cancel each other. In fact, the only terms that give nonzero contributions to the constraint algebra are those that contain the spatial derivative $h'_{ij}$ of the metric in one argument of the Poisson bracket and the conjugate momentum $\pi^{lm}$ in the other. This happens because only in that case do we get the derivative of a delta function, which prevents the term from being cancelled by the identical Poisson bracket where the two functionals are exchanged. Then, taking into account that the boundary conditions are chosen such that the constraints vanish at infinity, it is possible to shift these derivatives to $N$ and $M$ thanks to an integration by parts. Following these steps, one can work out the Dirac algebra. In the light of this, it is clear that the measure-dependent term of the Hamiltonian constraint with weighted derivatives does not affect the Poisson bracket $\{H[N],H[M]\}$. 

As a result, we can claim that standard diffeomorphism invariance is preserved in the multifractional theory with weighted derivatives in the absence of matter, since the $\phi$-dependent correction term is not affected by diffeomorphisms. When interacting matter fields are present, diffeomorphism invariance is broken \cite{frc11}. As far as LQG is concerned, the absence of deformations in the HDA excludes a relation between the theory with weighted derivative and the LQG formulation where anomaly freedom is imposed, while the differences in the cosmological dynamics \cite{frc11} exclude a connection also in LQG approaches with undeformed HDA.


\section{Conclusions}\label{concl}
 
In this paper, we have explored the similarities between $\kappa$-Minkowski and other noncommutative spacetimes with multifractional spacetimes. We found no exact duality between these two mutually disconnected regions of the landscape of multiscale theories. By making the multifractional theory with $q$-derivatives noncommutative via a canonical quantization of the geometric coordinates, we reproduced $\kappa$-Minkowski spacetime in the deep UV limit of the multiscale measure, in a much more general way than in \cite{ACOS}. All these results have been obtained at the level of the Heisenberg and Poincaré algebras of spacetime, i.e., by using symmetry arguments only. Making symmetry algebras central in the discussion is an efficient way to keep contact with phenomenology, since dispersion relations, the design of experiments involving the elementary measurements of lengths and times, and other aspects related to the physical testing of these theories can all be derived from the spacetime algebras considered here.

This study settles an issue left open in the literature. Having discovered that the conjectured duality is not present, we have shed light on the mutual relation between noncommutative and multifractional spacetimes. It is now clear that these occupy different places in the wide zoo of multiscale theories \cite{trtls} and, thus, they have to be considered as two independent and distinct approaches, which should be experimentally tested independently. The loss of a duality forbids to merge these two proposals and, in parallel, highlights their weak points. More attention should be paid to dimensional flow in noncommutative spacetimes, while nonfactorizable measures \cite{fra1,fra2,fra3} deserve further investigations albeit only as phenomenological models \cite{revmu}. We feel confident that forthcoming efforts will polarize also into these expanding fields.

The findings of Sec.\ \ref{4} close a theoretical gap in the analysis of the gravitational dynamics in multifractional theories. We have computed the constraint algebra and compared its deformations, when present, with those of loop quantum gravity in the effective-dynamics approach. There was no reason, \emph{a priori}, to expect a perfect match between these deformations, partly because the multifractional cosmological dynamics \cite{frc11} is clearly different from that of loop quantum cosmology \cite{Boj08,lqcr1,lqcr2} and partly due to the mismatch, made clear in Sec.\ \ref{2}, between the symmetries of multifractional spacetimes and those of $\kappa$-Minkowski (compatible with the flat limit of LQG). However, the points of similarity with $\kappa$-Minkowski begged for further inspection in the context of the hypersurface-deformation algebra.

With respect to the two multifractional theories considered here, with weighted and $q$-derivatives, the theory with fractional derivatives differs only in the choice of kinetic terms \cite{frc2,frc1}. This choice does affect the structure of momentum space, so that both the deformed Heisenberg algebra and the deformed Poincaré symmetry algebra of the theory with fractional derivatives will most likely be different from the algebras we constructed above. Nevertheless, since everything said here is heavily conditioned by the factorizability property of the measure of multifractional theories, we expect all our general arguments to apply also to the case with fractional derivatives. In particular, this case should not be dual to any noncommutative theory and should admit a well-defined noncommutative extension giving rise to specific deformed symmetries. It will be interesting to verify these expectations, not only to complete the theoretical study of multifractional spacetimes but also because the theory with fractional derivatives may offer a viable framework where to quantize gravity perturbatively \cite{fra1,frc2}. We hope to report on that in the near future.

\section*{Acknowledgments}

G.C.\ is under a Ram\'on y Cajal contract and is supported by the I+D grant FIS2014-54800-C2-2-P. M.R.\ thanks the Gravitation and Cosmology group of IEM-CSIC for partial support during his stay in Madrid, where this project was initiated. We also thank Martin Bojowald for useful correspondence.

\end{document}